\documentclass[12pt]{article}

\usepackage{latexsym,amsmath,amssymb,theorem,epsfig} 

\DeclareFontFamily{OT1}{rsfs10}{} 
\DeclareFontShape{OT1}{rsfs10}{m}{n}{ <-> rsfs10 }{} 
\DeclareMathAlphabet{\mathscript}{OT1}{rsfs10}{m}{n} 

\numberwithin{equation}{section}

\newcommand{\ns}{\normalsize}

\theoremstyle{plain} 

{\theorembodyfont{\rmfamily} }

\begin{document}


\begin{titlepage}

\vspace{-10cm}

\title{
   {\LARGE Visible Branes with Negative Tension in Heterotic M-Theory} \\[1em] } 
\author{
   Ron Y. Donagi$^1$, Justin Khoury$^2$, Burt A. Ovrut$^{3,4}$\\
   Paul J. Steinhardt$^2$ and Neil Turok$^5$\\[0.5cm]
   {\ns $^1$Department of Mathematics, University of Pennsylvania} \\[-0.4em]
   {\ns Philadelphia, PA 19104--6395, USA}\\
   {\ns $^2$Joseph Henry Laboratories, Princeton University} 
   \\[-0.4em]
   {\ns Princeton, NJ 08544, USA}\\ 
   {\ns $^3$Department of Physics, University of Pennsylvania} 
   \\[-0.4em]
   {\ns Philadelphia, PA 19104-6396, USA}\\ 
   {\ns $^4$Department of Physics and Astronomy, Rutgers University} 
   \\[-0.4em]
   {\ns Piscataway, NJ 08855-0849, USA}\\ 
   {\ns $^5$DAMTP, CMS, Wilberforce Road} \\[-0.4em]
   {\ns Cambridge, CB3 OWA, U.K.}\\}
\date{}

\maketitle

\begin{abstract}

It is shown that there exist large classes of BPS vacua in heterotic $M$-theory
which have negative tension on the visible orbifold plane, positive tension on the
hidden plane and positive tension, physical five-branes in the bulk space.
Explicit examples of such vacua are presented. Furthermore, it is
demonstrated that the ratio, $\beta/|\alpha|$, of the bulk five-brane tension 
to the visible plane tension can, for several large classes of such vacua, be made
arbitrarily small. Hence, it is straightforward to find vacua 
with the properties required in the examples of the Ekpyrotic theory of 
cosmology - a visible brane with negative tension and $\beta/|\alpha|$ small. 
This contradicts recent claims in the literature.

 \end{abstract}

\thispagestyle{empty}

\end{titlepage}


\section{Introduction}

In a recent paper \cite{EU}, a new theory of the early universe, called the
Ekpyrotic Universe, was introduced. This theory attempts to resolve the 
cosmological horizon, flatness and monopoles problems and to generate a 
nearly scale-invariant spectrum of energy density perturbations 
without requiring any period of inflation. Although, in principle, 
Ekpyrotic theory applies to any brane world scenario \cite{BW1,BW1B,
BW2,BW3,BW4,BW5,BW6},
in \cite{EU} it was specifically demonstrated in two concrete examples, $AdS$
spaces \cite{BW5} and heterotic $M$-theory \cite{BW1,BW1B,HW}.
In both cases, it was assumed
that the tension of the visible brane, that is, the brane supporting our
observable universe, is negative, whereas the tension of the hidden brane, that
is, the brane communicating with our world only through the extra dimension, 
is positive. Although the 
Ekpyrotic scenario does not, necessarily, require such an assumption, 
it was helpful in constructing the theories presented in \cite{EU}. It was further
assumed, fundamentally, that such theories contain physical branes in the bulk
space. In heterotic $M$-theory, where such questions can be analyzed
in detail, it is clear that these assumptions are valid for a wide class of 
physically relevant vacua. In this paper, we will discuss and explicitly construct
large classes of BPS vacua in heterotic $M$-theory that have negative tension visible
branes, positive tension hidden branes and physical, positive tension 
five-branes in the bulk space. These five-branes are wrapped on holomorphic
curves in the Calabi-Yau three-fold background.

In Ekpyrotic theory, the ratio of the bulk five-brane tension, $\beta$, 
to the magnitude of the tension, $|\alpha|$, of the visible brane arises as an 
important parameter. Although the Ekpyrotic scenario is potentially 
applicable for any value of this ratio, the theory is easiest to analyze when
$\beta/|\alpha|$ is small. In \cite{EU}, a sample value of this ratio was chosen to
be of order $10^{-4}$. In the same paper, it was demonstrated in a second
example that
values of $\beta/|\alpha|$ as large as $1/10$ are also acceptable. Be this as
it may, it is useful to demonstrate that small values of this ratio can be
obtained in brane world scenarios, so that we understand how much 
flexibility we
have in designing models. In this paper, we show explicitly
that arbitrarily small values of the ratio $\beta/|\alpha|$ can be obtained in
physically relevant BPS vacua of heterotic $M$-theory. These vacua have negative
tension visible branes, positive tension hidden branes and bulk space
five-branes with positive tension.

Specifically, we do the following. In Section 2, we present the relevant
structure of effective five-dimensional heterotic $M$-theory without bulk
five-branes. This is extended to include bulk five-branes in Section 3. We
define the tension of each orbifold boundary plane and bulk
five-brane in terms of characteristic classes. We demonstrate that, for
generic stable holomorphic vector bundles associated 
with the orbifold fixed planes,
the tension of the visible plane can be negative while the tension of the
hidden plane is positive. The tension of the bulk brane must always be
positive since it corresponds to an effective class, that is, it is wrapped on
a holomorphic curve of positive volume. In Section 4, we review
the formalism for computing the second Chern class of the tangent
bundle of elliptically fibered Calabi-Yau three-folds with either 
del Pezzo or Hirzebruch base surfaces, and the Chern classes for stable,
holomorphic vector bundles over such manifolds. This formalism is applied in
Section 5, where we present two explicit heterotic $M$-theory BPS vacua
which manifestly exhibit negative tension on the visible plane and 
positive tension on the hidden plane  and bulk branes. In both of these examples,
the visible sector supports an $SU(5)$ grand unified theory with three families
of quarks and leptons and the hidden sector an $E_{7}$ gauge group. In
Section 6, by discussing the relative volume of effective curves, we show that
there is a large class of $M$-theory BPS vacua whose branes exhibit the desired 
tensions, that is, negative and positive tensions on the visible and
hidden branes respectively, but not manifestly.
We present two explicit vacua of this type. The first has the same gauge
group and matter structure as previously. The second example has
an $SU(5)$ grand unified theory with three families of quarks and leptons 
on the visible plane with an $E_{6}$ gauge group on the hidden plane. In Section 7, 
we compute the ratio, $\beta/|\alpha|$, of the bulk
brane tension to the magnitude of the visible plane tension for each of the
four explicit $M$-theory vacua presented in this paper. While for two of 
these vacua this ratio is of order unity, we show that for the remaining two 
classes of vacua, this ratio can be made arbitrarily small.
In the conclusion, Section 8, we argue that our results can be extended to 
other elliptically fibered
Calabi-Yau three-folds, as well as other choices of gauge groups.
Finally, in an Appendix, we present a proof that the
volume of effective curves associated with the base can be made arbitrarily
large with respect to the volume of the elliptic fiber. This fact is essential
in the discussions in Sections 6 and 7. 

In a recent paper by Kallosh, Kofman and Linde \cite{KKL}, it was suggested that 
it is not possible to have negative
tension visible branes and positive tension hidden branes in heterotic
$M$-theory. This is incorrect. We show in Section 3 of this paper that it is
clear,
by construction, that heterotic $M$-theory vacua with non-standard gauge embeddings
(that is, general stable, holomorphic vector bundles) can have the
tensions as stated in \cite{EU}. Furthermore, it is very simple
to create large classes of such vacua, as is done in Sections 5 and 6 of
this paper. The authors
of \cite{KKL} refer primarily to the so-called standard embeddings. 
These embeddings, indeed, have
the tensions reversed. However, these vacua are long since known not to allow
five-branes in the bulk space and, hence, are clearly not appropriate for the 
Ekpyrotic scenario \cite{EU}. In their revised version, the authors of
\cite{KKL} claim that even for non-standard embeddings it is not
possible to get a negative tension visible brane and a positive tension hidden
brane. They state as the reason that the gauge group on the visible sector
must be smaller than the gauge group on the hidden sector. They conclude from
this that the tension on the visible brane, therefore, must be larger than the
tension of the hidden brane. This also is not correct, as all the examples in
this paper will clearly demonstrate. Quite to the contrary, 
we show that it is straightforward 
to construct vacua in heterotic M-theory satisfying the requirements of 
examples of Ekpyrotic cosmology presented in Ref.~\cite{EU}. 

\section{Effective Theory Without Bulk Five-Branes}

As first shown in \cite{HW}, cancellation of both gravitational and gauge anomalies
on the orbifold fixed planes requires that the Bianchi identity associated
with the three-form $C_{IJK}$, $I,J=0,\dots,9,11$, be modified to
\begin{equation}
(dG)_{11\bar{I}\bar{J}\bar{K}\bar{L}}=-4\sqrt{2}\pi\left(\frac{\kappa}
{4\pi}\right)^{2/3}(J^{(1)}\delta(x^{11}) + J^{(2)}\delta(x^{11}-\pi\rho))_{\bar{I}
\bar{J}\bar{K}\bar{L}},
\label{eq:1}
\end{equation}
where $\bar{I},\bar{J},\dots=0,\dots,9$ are ten-dimensional indices and $G=dC$
is the field strength. The sources are defined by
\begin{equation}
J^{(n)}= c_{2}(V^{n})-\frac{1}{2}c_{2}(TX) \;\;\;\;\;\; n=1,2,
\label{eq:2}
\end{equation}
and
\begin{equation}
c_{2}(V^{n})=\frac{1}{16\pi^{2}}trF^{n} \wedge F^{n}, \qquad
c_{2}(TX)=\frac{1}{16\pi^{2}}trR \wedge R,
\label{eq:3}
\end{equation}
where $F^{n}$ is the field strength associated with the gauge theory on the $n$-th plane, and $R$ is the Ricci tensor of the Calabi-Yau manifold.
Note that $c_{2}(V^{n})$ and $c_{2}(TX)$ are the second Chern class of the
vector bundle on the $n$-th boundary plane and the second Chern class of the
Calabi-Yau tangent bundle respectively. Integrating (\ref{eq:1}) over a
five-cycle which spans the orbifold interval and is otherwise an arbitrary
four-cycle in the Calabi-Yau three-fold, we find
\begin{equation}
c_{2}(V^{1})+c_{2}(V^{2})-c_{2}(TX)=0.
\label{eq:4}
\end{equation}
It follows that the sources must satisfy
\begin{equation}
J^{(1)}=-J^{(2)}.
\label{eq:5}
\end{equation}
Note that this implies that the bundles on the two fixed planes must, in
general, be different.

The effective five-dimensional theory with two four-dimensional
boundaries, but without bulk five-branes, was computed in \cite{BW1B} 
and was found to be
\begin{equation}
S_{5}=S_{bulk}+S_{boundary},
\label{eq:6}
\end{equation}
where
\begin{equation}
S_{bulk}=-\frac{1}{2\kappa_{5}^{2}}\int_{M_{5}}{\sqrt{-g}(R+\frac{1}{2}
V^{-2}\partial_{\alpha}V\partial^{\alpha}V+\cdots)}
\label{eq:7}
\end{equation}
and
\begin{equation}
S_{boundary}=-\frac{\sqrt{2}}{\kappa_{5}^{2}}\int_{M_{4}^{(1)}}{\sqrt{-g}V^{-1}
\alpha_{i}^{(1)}b^{i}} - \frac{\sqrt{2}}{\kappa_{5}^{2}}\int_{M_{4}^{(2)}}
{\sqrt{-g}V^{-1}\alpha_{i}^{(2)}b^{i}}.
\label{eq:8}
\end{equation}
Note the we have explicitly written only those terms in (\ref{eq:7}) and
(\ref{eq:8}) that 
are relevant to this discussion. The various quantities that make up this
effective theory are defined as follows. First,
\begin{equation}
\kappa_{5}^{2}=\frac{\kappa^{2}}{v}
\label{eq:9}
\end{equation}
where $\kappa$ is the 11-dimensional gravitational coupling constant and $v$ is a
6-dimensional reference volume. The moduli fields arise from the structure of
the 11-dimensional line element, which has the form
\begin{equation}
ds^{2}=V^{-2/3}g_{\alpha\beta}dx^{\alpha}dx^{\beta}+g_{AB}dx^{A}dx^{B}
\label{eq:10}
\end{equation}
The indices take values $\alpha,\beta=0,1,2,3,11$ and $A,B=4,\dots,9$ 
with the
latter corresponding to the six real 
coordinates of the Calabi-Yau three-fold. Then
\begin{equation}
V=\frac{1}{v}\int_{X}{\sqrt{^{6}g}}
\label{eq:11}
\end{equation}
where $^{6}g$ is the determinant of $g_{AB}$. Now define metric
\begin{equation}
\Omega_{AB}=V^{-1/3}g_{AB}
\label{eq:12}
\end{equation}
and let $\omega_{a \bar{b}}=i\Omega_{a \bar{b}}$ be its K\"{a}hler form, where
$a,b$ are the complex coordinates corresponding to $A,B$.
Choosing a basis $\omega_{iAB}, i=1,\dots,h^{1,1}$ of harmonic (1,1)-forms 
of $H^{(1,1)}(X)$, whose Poincar\'e duals, ${\cal{C}}_{i}$, 
are cycles in $H_{4}(X,{\bf Z \rm})$, we can expand
\begin{equation}
\omega_{AB}= b^{i}\omega_{iAB}
\label{eq:13}
\end{equation}
which defines the $h^{1,1}$ moduli $b^{i}$. Finally, the coefficients
$\alpha_{i}^{(n)}$ are given by
\begin{equation}
\alpha_{i}^{(n)}=\frac{2\sqrt{2}\pi}{v^{2/3}}\left(\frac{\kappa}{4\pi}\right)^{2/3}
\int_{{\cal{C}}_{i}}{J^{(n)}}.
\label{eq:14}
\end{equation}
The $h^{1,1}=h_{4}$ four-cycles ${\cal{C}}_{i}$ span 
$H_{4}(X,{\bf Z \rm})$ and, by definition, satisfy
\begin{equation}
\int_{{\cal{C}}_{i}}{\nu}=\int_{X}{\omega_{i}\wedge \nu}
\label{eq:15}
\end{equation}
for any four-form $\nu$. Note from equations (\ref{eq:5}) and (\ref{eq:14})
that
\begin{equation}
\alpha_{i}^{(1)}=-\alpha_{i}^{(2)}
\label{eq:16}
\end{equation}
for any value of $i$. 

For arbitrary four-form $\nu$,
\begin{equation}
b^{i}\int_{{\cal{C}}_{i}}{\nu}=\int_{b^{i}{\cal{C}}_{i}}{\nu}=\int_{X}{b^{i}
\omega_{i}\wedge \nu}.
\label{eq:17}
\end{equation}
It then follows from (\ref{eq:13}) and (\ref{eq:15}) that
\begin{equation}
b^{i}\int_{{\cal{C}}_{i}}{\nu}=\int_{{\cal{C}}_{\omega}}{\nu},
\label{eq:18}
\end{equation}
where ${\cal{C}}_{\omega}$ is the Poincar\'e dual four-cycle to the K\"{a}hler
class $\omega$. We conclude that
\begin{equation}
\alpha_{i}^{(n)}b^{i}=\alpha^{(n)},
\label{eq:19}
\end{equation}
where
\begin{equation}
\alpha^{(n)}=\frac{2\sqrt{2}\pi}{v^{2/3}}\left(\frac{\kappa}{4\pi}\right)^{2/3}
\int_{{\cal{C}}_{\omega}}{J^{(n)}}.
\label{eq:20}
\end{equation}
Again, one finds from (\ref{eq:5}) that
\begin{equation}
\alpha^{(1)}=-\alpha^{(2)}.
\label{eq:21}
\end{equation}
We conclude that the boundary part of the action is given by
\begin{equation}
S_{boundary}=-\frac{\sqrt{2}}{\kappa_{5}^{2}}\int_{M_{4}^{(1)}}{\sqrt{-g}V^{-1}
\alpha^{(1)}} - \frac{\sqrt{2}}{\kappa_{5}^{2}}\int_{M_{4}^{(2)}}
{\sqrt{-g}V^{-1}\alpha^{(2)}},
\label{eq:22}
\end{equation}
where, using (\ref{eq:2}) and (\ref{eq:20}) we have
\begin{equation}
\alpha^{(n)}=\frac{2\sqrt{2}\pi}{v^{2/3}}\left(\frac{\kappa}{4\pi}\right)^{2/3}
\int_{{\cal{C}}_{\omega}}{(c_{2}(V^{n})-\frac{1}{2}c_{2}(TX))}.
\label{eq:23}
\end{equation}
Note that, on the boundaries, the K\"{a}hler form $\omega$ and 
moduli $b^{i}$ are generically functions of $x^{\mu}$, $\mu=0,1,2,3$ and,
hence, so are $\alpha^{(n)}$. However, in this paper,
we are interested in the brane tensions associated with the universal BPS
vacuum solutions discussed in \cite{BW1B}. 
In these solutions, the $b^{i}$ moduli are
constants. It follows that, for these vacua, the K\"{a}hler form and, therefore, 
$\alpha^{(n)}$ are constants. We will assume this for the remainder of this
paper.

From the form of the action given by (\ref{eq:6}),(\ref{eq:7}) and
(\ref{eq:22}), we see that the coefficient $\alpha^{(n)}$ defines the tension on the
$n$-th orbifold plane. Specifically, if $\alpha^{(n)}>0$ then the tension of
the associated boundary plane is positive, whereas if $\alpha^{(n)}<0$ the
tension on that orbifold plane is negative. It is clear from the expression
for $\alpha^{(n)}$ in (\ref{eq:23}) that its value
and sign depend on the explicit choice of the Calabi-Yau three-fold (thus
specifying $c_{2}(TX)$) and on the holomorphic vector bundle $V^{n}$ (thus
specifying $c_{2}(V^{n})$).

The simplest example one can present is the so-called 
standard embedding, where one fixes a Calabi-Yau three-fold and chooses
the two holomorphic vector bundles so that $V^{1}=TX$ and $V^{2}=0$. It
follows that
\begin{equation}
c_{2}(V^{1})=c_{2}(TX), \qquad c_{2}(V^{2})=0.
\label{eq:24}
\end{equation}
Note that these Chern classes satisfy the topological condition given
in (\ref{eq:4}). From (\ref{eq:23}) we find
\begin{equation}
\alpha^{(1)}=\frac{\sqrt{2}\pi}{v^{2/3}}\left(\frac{\kappa}{4\pi}\right)^{2/3}
\int_{{\cal{C}}_{\omega}}{c_{2}(TX)}
\label{eq:25}
\end{equation}
which is positive. Hence, the tension on the $1$-plane is positive.
Similarly, the coefficient $\alpha^{(2)}$ is negative, consistent
with expression (\ref{eq:21}), and, hence, the tension on the $2$-plane is
negative. The standard embedding can be given a particle physics
interpretation. Note that the bundle $V^{1}=TX$ must have
structure group $G=SU(3)$, whereas bundle $V^{2}$ is zero. It
follows that the surviving low-energy gauge symmetry on the $1$-plane is
$H=E_{6}$ (the commutant of $SU(3)$ in $E_{8}$). The gauge symmetry on the
$2$-plane remains $E_{8}$. Since $E_{6}$ might be interpreted as a viable
grand unification group, it is conventional to call the $1$-plane the
``visible'' or ``observable'' sector and the $2$-plane the ``hidden '' sector.
With these definitions, we see that  -in the standard embedding- 
the visible sector has positive tension and the hidden sector negative
tension.

Although they are well-known, bundles satisfying the standard embedding
conditions suffer from a number of difficulties. To begin with, as
we have said, they lead to an $E_{6}$ grand unified theory (GUT) in the
visible sector which, having a large rank and dimension, is more difficult to 
break to the standard model group $SU(3)_{C} \times SU(2)_{L} \times U(1)_{Y}$.
In addition, although it can be done, it is hard to obtain a spectrum 
with three light families of 
quarks and leptons. Furthermore, even though the standard embedding is natural 
in the context of $E_{8} \times E_{8}$ heterotic superstring theory, there is
no reason for preferring it in $M$-theory, where all consistent stable, 
holomorphic vector bundle vacua are
on an equal footing. Finally, as we will show explicitly in the next section,
since the standard embedding satisfies the topological condition (\ref{eq:4}),
it does not admit any five-branes in the five-dimensional bulk space. Therefore,
the standard embedding cannot be used in the recently proposed Ekpyrotic 
cosmological theory \cite{EU}, or, for that matter, in the variant given in 
\cite{KKL}.

For all these reasons, it is of interest to construct ``non-standard
embeddings'', that is, vacua in which the spin connection of the Calabi-Yau 
three-fold is not embedded in the gauge group. For vacua without bulk
five-branes, such non-standard embeddings are described by stable, holomorphic
vector bundles satisfying the topological condition (\ref{eq:4}), but not
(\ref{eq:24}). One property of these vacua is immediately apparent. By
choosing vector bundles $V^{1}$ and $V^{2}$ so that
\begin{equation}
\int_{{\cal{C}}_{\omega}}{c_{2}(V^{1})} < \frac{1}{2}\int_{{\cal{C}}_{\omega}}
{c_{2}(TX)},
\label{eq:26}
\end{equation}
one can make $\alpha^{(1)}<0$ and, using (\ref{eq:21}), $\alpha^{(2)}>0$. As we
will show by explicit example in the next section, vector bundles
$V^{1}$ satisfying (\ref{eq:26}) can break $E_{8}$ in the visible sector 
down to small gauge groups, such as $SU(5)$. The examples in this paper are
easily extended to other small gauge groups, such as $SO(10)$ and the standard model 
gauge group $SU(3)_{C} \times SU(2)_{L} \times U(1)_{Y}$. It follows,
therefore, that non-standard embeddings can easily produce vacua with negative
tension on the visible orbifold plane and positive tension on the hidden
sector fixed plane. It is not too difficult to produce examples of
non-standard embeddings satisfying (\ref{eq:4}) and (\ref{eq:26}). However,
such vacua continue to suffer from the last difficulty mentioned above, namely
that, since they do not  admit bulk five-branes, they cannot be
used as vacua for the Ekpyrotic cosmological theory. We turn, therefore, to
more generalized vacua of $M$-theory that do admit bulk five-branes.

\section{Effective Theory With Bulk Five-Branes}

When $N$ bulk five-branes, located at coordinates $x_{i}$ for $i=1,\dots,N$
in the 11-direction, are present in the vacuum, cancellation of their
worldvolume anomalies, as well as the gravitational and gauge anomalies 
on the orbifold fixed planes, requires that Bianchi identity (\ref{eq:1}) be  
further modified to
\begin{equation}
(dG)_{11\bar{I}\bar{J}\bar{K}\bar{L}}=-4\sqrt{2}\pi\left(\frac{\kappa}
{4\pi}\right)^{2/3}(J^{(1)}\delta(x^{11}) + J^{(2)}\delta(x^{11}-\pi\rho)
+\Sigma_{i=1}^{N} \hat{J}^{(i)}\delta(x^{11}-x_{i}))_{\bar{I}
\bar{J}\bar{K}\bar{L}}.
\label{eq:27}
\end{equation}
Each five-brane source $\hat{J}^{(i)}$ is defined to be the four-form which is 
Poincar\'e dual to the holomorphic curve in the Calabi-Yau three-fold 
around which the $i$-th five-brane is wrapped. If we define the five-brane class 
\begin{equation}
W=\Sigma_{i=1}^{N} \hat{J}^{(i)},
\label{eq:28}
\end{equation}
then the topological condition (\ref{eq:4}) is modified to
\begin{equation}
c_{2}(V^{1})+c_{2}(V^{2})-c_{2}(TX)+W=0.
\label{eq:29}
\end{equation}
As we will see, the addition of bulk five-branes makes solution of this
topological condition considerably simpler.

The effective five-dimensional theory with two four-dimensional boundaries and
$N$ bulk five-branes was discussed in \cite{NSE}. 
For the remainder of this paper, we
will assume, for simplicity, that all bulk five-branes are located at the same
point in the orbifold direction. This assumption corresponds 
to choosing a specific region of
the moduli space of $W$ and is sufficient for our purposes. 
A detailed analysis of the full moduli space of five-brane classes $W$ was 
presented in \cite{FB}. None of the conclusions arrived 
at in this section will change
if one considers other regions of moduli space. With this simplifying
assumption, the effective theory is found to be
\begin{equation}
S_{5}=S_{bulk}+S_{boundary}+S_{5-brane},
\label{eq:30}
\end{equation}
where the first few terms of $S_{bulk}$ are given in (\ref{eq:7}). The
boundary and five-brane actions are 
\begin{equation}
S_{boundary}=-\frac{\sqrt{2}}{\kappa_{5}^{2}}\int_{M_{4}^{(1)}}
{\sqrt{-g}V^{-1}\alpha^{(1)}} - \frac{\sqrt{2}}{\kappa_{5}^{2}}\int_{M_{4}^{(2)}}
{\sqrt{-g}V^{-1}\alpha^{(2)}} 
\label{eq:31}
\end{equation}
where
\begin{equation}
\alpha^{(n)}=\frac{2\sqrt{2}\pi}{v^{2/3}}\left(\frac{\kappa}{4\pi}\right)^{2/3}
\int_{{\cal{C}}_{\omega}}{(c_{2}(V^{n})-\frac{1}{2}c_{2}(TX))},
\label{eq:32}
\end{equation}
and
\begin{equation}
S_{5-brane}=-\frac{\sqrt{2}}{\kappa_{5}^{2}}\int_{M_{4}^{(5-brane)}}
{\sqrt{-g}V^{-1}\beta},
\label{eq:33}
\end{equation}
with
\begin{equation}
\beta=\frac{2\sqrt{2}\pi}{v^{2/3}}\left(\frac{\kappa}{4\pi}\right)^{2/3}
\int_{{\cal{C}}_{\omega}}{W}.
\label{eq:34}
\end{equation}
Note from topological condition (\ref{eq:29}) that
\begin{equation}
\alpha^{(1)}+\alpha^{(2)}+\beta=0.
\label{eq:35}
\end{equation}

Before considering non-standard embeddings, let us briefly discuss how the
standard embedding, that is, holomorphic vector bundles satisfying 
(\ref{eq:24}), fit into this context. For the standard embedding, it follows
from (\ref{eq:32}) that $\alpha^{(1)}=-\alpha^{(2)}$ and, hence, from
(\ref{eq:35}) that $\beta=0$. That is, the standard embedding always has
vanishing five-brane class $W$ and, therefore, no five-branes in the bulk, as
anticipated in the previous section. We now proceed to the construction of
non-standard embedding vacua. To do this, we need to introduce specific
Calabi-Yau three-folds and give a method for constructing stable,
holomorphic vector bundles over them. Once this is accomplished, we can use
expressions (\ref{eq:32}) and (\ref{eq:34}) to calculate the tension of the visible
and hidden orbifold planes, as well as the bulk five-brane.

\section{Vector Bundles and Chern Classes}

In this paper, we will consider Calabi-Yau three-folds that are elliptically
fibered over a base surface. The stable, holomorphic vector
bundles over these Calabi-Yau three-folds will be constructed using the methods
introduced in \cite{FMW1,AJ,FMW2} and \cite{RD1,RD2,RD3,RD4,C}. 
Here, we only state the results that we
specifically need to formulate realistic particle physics models and to
compute their brane tensions. We refer the reader to 
\cite{RD1,RD2,RD3,RD4} where these topics are discussed in detail.

The second Chern class of an elliptically fibered Calabi-Yau three-fold
$X$ over a base surface $B$ is given by
\begin{equation}
c_{2}(TX)= c_{2}(B)+11c_{1}(B)^{2}+12 \sigma c_{1}(B),
\label{eq:36}
\end{equation}
where $c_{1}(B)$ and $c_{2}(B)$ are the first and second Chern classes of
the base manifold respectively and $\sigma$ is the zero section of $X$. 
$B$ is restricted to be either a
del Pezzo, Hirzebruch or Enriques surface, or a blow-up of a Hirzebruch
surface. The Chern classes are known for all of these. In this paper
we will consider, for specificity, del Pezzo and Hirzebruch surfaces only. 
The other allowed bases can also be considered. These have the
following Chern classes. For del Pezzo surfaces $dP_{r}$, $r=1,\dots,8$, the
Chern classes are given by
\begin{equation}
c_{1}(dP_{r})=3l-\Sigma_{i=1}^{r}E_{i}, \qquad c_{2}=3+r,
\label{eq:37}
\end{equation}
where $l$ and $E_{i}$, $i=1,\dots,r$ are effective curves spanning
$H_{2}(dP_{r})$. Their intersection numbers are
\begin{equation}
l \cdot l=1, \qquad l \cdot E_{i}=0, \qquad E_{i} \cdot E_{j}=-\delta_{ij}.
\label{eq:38}
\end{equation}
For Hirzebruch surfaces $F_{r}$, $r$ any non-negative integer, the Chern
classes are given by
\begin{equation}
c_{1}(F_{r})=2S+(r+2){\cal{E}}, \qquad c_{2}(F_{r}) =4,
\label{eq:39}
\end{equation}
where $S$ and $\cal{E}$ are effective curves spanning $H_{2}(F_{r})$. Their
intersection numbers are
\begin{equation}
S \cdot S=-r, \qquad S \cdot {\cal{E}}=1, \qquad {\cal{E}} \cdot {\cal{E}}=0.
\label{eq:40}
\end{equation}

The Chern classes for a family of stable, holomorphic vector bundles $V$
over elliptically fibered Calabi-Yau three-folds were computed in
\cite{FMW1,FMW2}. For
structure group $G=SU(n)$, they were found to be
\begin{equation}
c_{2}(V)=\sigma \eta - \frac{1}{24}c_{1}(B)^{2}(n^{3}-n) + \frac{1}{2}
\left(\lambda^{2}-\frac{1}{4}\right)n\eta(\eta-nc_{1}(B))
\label{eq:41}
\end{equation}
and
\begin{equation}
c_{3}(V)=2\lambda \sigma \eta(\eta-nc_{1}(B)).
\label{eq:42}
\end{equation}
Here $\eta$ is any effective curve in the base surface $B$ and $\lambda$ is a
rational number subject to the constraints
\begin{equation}
n \quad \mbox{is odd}, \qquad \lambda=m+\frac{1}{2}
\label{eq:43}
\end{equation}
or
\begin{equation}
n \quad \mbox{is even}, \qquad \lambda=m, \quad \eta=c_{1}(B)mod2,
\label{eq:44}
\end{equation}
where $m$ is an integer.

It follows from (\ref{eq:29}) that
\begin{equation}
W=c_{2}(TX)-c_{2}(V^{1})-c_{2}(V^{2}).
\label{eq:45}
\end{equation}
Using expressions (\ref{eq:36}) and (\ref{eq:41}), we can write the
five-brane class $W$ as
\begin{equation}
W=W_{B}+a_{f}F,
\label{eq:46}
\end{equation}
where
\begin{equation}
W_{B}=\sigma(12c_{1}(B)-\eta^{(1)}-\eta^{(2)})
\label{eq:47}
\end{equation}
is the component of the class associated with the base $B$ and
\begin{eqnarray}
\nonumber
& & a_{f}= c_{2}(B)+\left(11 +\frac{n^{(1)3}-n^{(1)}}{24} +\frac{n^{(2)3}-n^{(2)}}{24}\right)c_{1}(B)^{2} \\
\nonumber
& & \;\;\;\;\;\;\;\;-\frac{1}{2}n^{(1)}\left(\lambda^{(1)2}-\frac{1}{4}\right)\eta^{(1)}(\eta^{(1)}-n^{(1)}c_{1}(B)) \\
& & \;\;\;\;\;\;\;\;-\frac{1}{2}n^{(2)}\left(\lambda^{(2)2}-\frac{1}{4}\right)\eta^{(2)}(\eta^{(2)}-n^{(2)}c_{1}(B))
\label{eq:48}
\end{eqnarray}
is the integer coefficient associated with the elliptic fiber class $F$. Here, the
superscript $(i)$, $i=1,2$ refers to the stable, holomorphic vector bundle $V^{i}$
on the $i$-th orbifold plane. As discussed in \cite{RD1}, there is an important
additional constraint that five-brane class $W$ must satisfy. That is, 
in order for $W$ to represent a set of physical five-branes in the bulk 
it must be an effective class. It was shown in \cite{RD1} that this will be 
the case if and only if
\begin{equation}
W_{B} \quad \mbox{is effective in B}, \quad a_{f}\geq 0.
\label{eq:49}
\end{equation}
This condition strongly constrains the allowed vacua.
The results presented in this section allow one to construct realistic vacua
of $M$-theory, to which we now turn.

\section{Physical Vacua and Brane Tension}

There are many  physically realistic $M$-theory vacua that will
have the three properties that we wish to explore, namely, negative tension on
the visible orbifold plane, positive tension on the hidden plane and
non-vanishing five-branes in the bulk. To simplify our discussion, in this
paper we will assume that $V^{1}$, the stable, holomorphic vector bundle on
the $1$-plane, has structure group 
\begin{equation}
G^{(1)}=SU(5).
\label{eq:50}
\end{equation}
That is, we choose
\begin{equation}
n^{(1)}=5.
\label{eq:51}
\end{equation}
It follows that the low energy gauge group on the $1$-plane is
\begin{equation}
H^{(1)}=SU(5)
\label{eq:52}
\end{equation}
which is the commutant subgroup of $G^{(1)}=SU(5)$ in $E_{8}$.
Furthermore, it was shown in \cite{C} that the number 
of quark and lepton families is given by
\begin{equation}
N_{family}=\frac{c_{3}(V)}{2}.
\label{eq:53}
\end{equation}
We, henceforth, demand that the number of families on the $1$-plane be the
observed number, that is, three. It follows from (\ref{eq:42}) that
\begin{equation}
\lambda^{(1)} \sigma \eta^{(1)}(\eta^{(1)}-5c_{1}(B))=3.
\label{eq:54}
\end{equation}
Conditions (\ref{eq:51}) and (\ref{eq:54}) imply that the $1$-plane supports 
an SU(5) GUT with three families of quarks and leptons. That is, these conditions
lead to a physically realistic particle model on the $1$-plane and, hence, this
plane is the visible sector. We find from (\ref{eq:32}), (\ref{eq:36}) 
and (\ref{eq:41}) that the tension of the visible sector is
\begin{equation}
\alpha^{(1)}=\frac{2\sqrt{2}\pi}{v^{2/3}}\left(\frac{\kappa}{4\pi}\right)^{2/3}
\int_{{\cal{C}}_{\omega}}{(c_{2}(V^{1})-\frac{1}{2}c_{2}(TX))}
\label{eq:55}
\end{equation}
where
\begin{equation}
c_{2}(V^{1})-\frac{1}{2}c_{2}(TX)=(\eta^{(1)}-6c_{1}(B))\sigma
-\frac{1}{2}\left(c_{2}(B)+21c_{1}(B)^{2}-\frac{15}{\lambda^{(1)}}
\left(\lambda^{(1)2}-\frac{1}{4}\right)\right)F
\label{eq:56}
\end{equation}
and we have used (\ref{eq:51}) and (\ref{eq:54}). Note that since we have
chosen $n^{(1)}$ to be odd, it follows from (\ref{eq:43}) that
\begin{equation}
\lambda^{(1)}=m^{(1)}+\frac{1}{2}
\label{eq:57}
\end{equation}
where $m^{(1)}$ is any integer and $\eta^{(1)}$ must be effective in $B$ but is
otherwise unconstrained.

Similarly, to simplify the discussion, we will in this section assume that
$V^{2}$, the stable, holomorphic vector bundle on the $2$-plane, has
structure group
\begin{equation}
G^{(2)}=SU(2).
\label{eq:58}
\end{equation}
That is, we choose
\begin{equation}
n^{(2)}=2.
\label{eq:59}
\end{equation}
It follows that the low energy gauge group on the $2$-plane is
\begin{equation}
H^{(2)}= E_{7}
\label{eq:60}
\end{equation}
which is the commutant subgroup of $G^{(2)}=SU(2)$ in $E_{8}$. We need impose
no further restrictions. Condition (\ref{eq:59}) implies that
the $2$-plane supports an $E_{7}$ supergauge theory with a certain number of
matter multiplets that we need not specify. That is, the $2$-plane is typical
of a hidden sector. We find from (\ref{eq:32}), (\ref{eq:36}) 
and (\ref{eq:41}) that the tension of the hidden sector is
\begin{equation}
\alpha^{(2)}=\frac{2\sqrt{2}\pi}{v^{2/3}}\left(\frac{\kappa}{4\pi}\right)^{2/3}
\int_{{\cal{C}}_{\omega}}{(c_{2}(V^{2})-\frac{1}{2}c_{2}(TX))}
\label{eq:61}
\end{equation}
where
\begin{eqnarray}
\nonumber
& & c_{2}(V^{2})-\frac{1}{2}c_{2}(TX)=(\eta^{(2)}-6c_{1}(B))\sigma
-\frac{1}{2}(c_{2}(B)+\frac{23}{2}c_{1}(B)^{2} \\
& & \;\;\;\;\;\;\;\;\;\;\;\;-2\left(\lambda^{(2)2}-\frac{1}{4}\right)\eta^{(2)}(\eta^{(2)}-2c_{1}(B))F
\label{eq:62}
\end{eqnarray}
and we have used (\ref{eq:59}). Note from (\ref{eq:44}) that, 
since we have chosen $n^{(2)}$ to be even, we must have
\begin{equation}
\lambda^{(2)}=m^{(2)}
\label{eq:63}
\end{equation}
where $m^{(2)}$ is any integer. Furthermore, in addition to being effective in
$B$, $\eta^{(2)}$ must satisfy
\begin{equation}
\eta^{(2)}=c_{1}(B) mod2.
\label{eq:64}
\end{equation}

Finally, using the conditions (\ref{eq:51}), (\ref{eq:54}) and (\ref{eq:59}),
the expressions for the five-brane class given in (\ref{eq:47}) and
(\ref{eq:48}) simplify to
\begin{equation}
W_{B}=\sigma(12c_{1}(B)-\eta^{(1)}-\eta^{(2)})
\label{eq:65}
\end{equation}
and
\begin{eqnarray}
\nonumber
& & a_{f}= c_{2}(B)+\left(\frac{65}{4}\right)c_{1}(B)^{2}
-\frac{15}{2\lambda^{(1)}}\left(\lambda^{(1)2}-\frac{1}{4}\right) \\
& & \;\;\;\;\;\;\;\;-\left(\lambda^{(2)2}-\frac{1}{4}\right)\eta^{(2)}(\eta^{(2)}-2c_{1}(B)).
\label{eq:66}
\end{eqnarray}
In terms of these quantities, the tension on the five-brane is given by
\begin{equation}
\beta=\frac{2\sqrt{2}\pi}{v^{2/3}}\left(\frac{\kappa}{4\pi}\right)^{2/3}
\int_{{\cal{C}}_{\omega}}{W}
\label{eq:67}
\end{equation}

We now present several heterotic $M$-theory vacua within this context that
explicitly have negative tension on the visible $1$-plane, positive tension on
the hidden $2$-plane and physical five-branes in the bulk with positive brane
tension.

\section*{Example 1: $dP_{8}$}

In this example, we will choose the base surface of the elliptically fibered
Calabi-Yau three-fold to be
\begin{equation}
B=dP_{8}.
\label{eq:68}
\end{equation}
Using (\ref{eq:37}), (\ref{eq:38}) and the fact that $c_{1}(dP_{r})^{2}=9-r$, it
follows that, for $r=8$, Eq.~(\ref{eq:56}) becomes 
\begin{equation}
c_{2}(V^{1})-\frac{1}{2}c_{2}(TX)=(\eta^{(1)}-6c_{1}(B))\sigma
-\frac{1}{2}\left(32-\frac{15}{\lambda^{(1)}}
\left(\lambda^{(1)2}-\frac{1}{4}\right)\right)F,
\label{eq:69}
\end{equation}
Eq.~(\ref{eq:62}) simplifies to
\begin{eqnarray}
\nonumber
& & c_{2}(V^{2})-\frac{1}{2}c_{2}(TX)=(\eta^{(2)}-6c_{1}(B))\sigma \\
& & -\frac{1}{2}\left(\frac{45}{2} -2\left(\lambda^{(2)2}-\frac{1}{4}\right)\eta^{(2)}(\eta^{(2)}-2c_{1}(B))\right)F,
\label{eq:70}
\end{eqnarray}
and the five-brane class specified by (\ref{eq:65}) and (\ref{eq:66}) becomes
\begin{equation}
W_{B}=\sigma(12c_{1}(B)-\eta^{(1)}-\eta^{(2)})
\label{eq:71}
\end{equation}
and
\begin{equation}
a_{f}= \frac{109}{4}
-\frac{15}{2\lambda^{(1)}}\left(\lambda^{(1)2}-\frac{1}{4}\right)
-\left(\lambda^{(2)2}-\frac{1}{4}\right)\eta^{(2)}(\eta^{(2)}-2c_{1}(B)),
\label{eq:72}
\end{equation}
where
\begin{equation}
c_{1}(dP_{8})=3l-\Sigma_{i=1}^{8}E_{i}.
\label{eq:73}
\end{equation}

We now choose
\begin{equation}
\lambda^{(1)}=-\frac{1}{2}, \qquad \eta^{(1)}=2c_{1}(dP_{8}).
\label{eq:74}
\end{equation}
Note that this value of $\lambda^{(1)}$ 
satisfies (\ref{eq:57}) and that
$\eta^{(1)}$ is effective since $c_{1}(dP_{8})$ is. It is straightforward to
show that these choices satisfy the three-family condition (\ref{eq:54}) on
the visible plane. Inserting this data into (\ref{eq:69}), we conclude that
\begin{equation}
c_{2}(V^{1})-\frac{1}{2}c_{2}(TX)=-4c_{1}(dP_{8})\sigma-16F.
\label{eq:75}
\end{equation}

Furthermore, let us take
\begin{equation}
\lambda^{(2)}=1, \qquad \eta^{(2)}=7c_{1}(dP_{8})
\label{eq:76}
\end{equation}
We see that this value of $\lambda^{(2)}$ satisfies (\ref{eq:63}) and that
$\eta^{(2)}=c_{1}(dP_{8})+6c_{1}(dP_{8})$ and, hence, satifies (\ref{eq:64}).
Clearly $\eta^{(2)}$ is effective since $c_{1}(dP_{8})$ is. Substituting this
data into (\ref{eq:70}), (\ref{eq:71}) and (\ref{eq:72}) then gives
\begin{equation}
c_{2}(V^{2})-\frac{1}{2}c_{2}(TX)=c_{1}(dP_{8})\sigma+15F
\label{eq:77}
\end{equation}
and
\begin{equation}
W=3c_{1}(dP_{8})\sigma+F,
\label{eq:78}
\end{equation}
respectively. Note that $W$ satisfies (\ref{eq:49}) and, hence, is an
effective class.

The tensions on the visible $1$-plane, the hidden
$2$-plane and the bulk five-brane are found by inserting these expressions in 
(\ref{eq:55}), (\ref{eq:61}) and (\ref{eq:67}) respectively. To evaluate the
signs of these tensions, it is necessary to recall that the integral of any 
effective four-form over the Poincar\'e dual of a K\"{a}hler class is simply the
volume of the complex curve associated with that four-form. Hence, this
integral is always positive. Specifically, it follows that
\begin{equation}
\int_{{\cal{C}}_{\omega}}{c_{1}(dP_{8})\sigma}>0, \qquad 
\int_{{\cal{C}}_{\omega}}{F}>0
\label{eq:79}
\end{equation}
since both $c_{1}(dP_{8})\sigma$ and $F$ are effective classes. Inserting
Eq.~(\ref{eq:75}) into (\ref{eq:55}), we find 
\begin{equation}
\alpha^{(1)}=-\frac{2\sqrt{2}\pi}{v^{2/3}}\left(\frac{\kappa}{4\pi}\right)^{2/3}
\int_{{\cal{C}}_{\omega}}{(4c_{1}(dP_{8})\sigma +16F)}
\label{eq:80}
\end{equation}
Using (\ref{eq:79}), we conclude that the tension on the visible 
$1$-plane is negative. Similarly, substituting Eq.~(\ref{eq:77}) 
into (\ref{eq:61}) yields
\begin{equation}
\alpha^{(2)}=\frac{2\sqrt{2}\pi}{v^{2/3}}\left(\frac{\kappa}{4\pi}\right)^{2/3}
\int_{{\cal{C}}_{\omega}}{(c_{1}(dP_{8})\sigma +15F)}
\label{eq:81}
\end{equation}
Hence, the tension on the hidden $2$-plane is positive. Finally,
putting Eq.~(\ref{eq:70}) into (\ref{eq:67}) gives
\begin{equation}
\beta=\frac{2\sqrt{2}\pi}{v^{2/3}}\left(\frac{\kappa}{4\pi}\right)^{2/3}
\int_{{\cal{C}}_{\omega}}{(3c_{1}(dP_{8})\sigma+F)}
\label{eq:82}
\end{equation}
Thus the five-brane has positive tension. As a final check on this
calculation, note that 
\begin{equation}
\alpha^{(1)}+\alpha^{(2)}+ \beta=0
\label{eq:83}
\end{equation}
as it must to satisfy the topological anomaly cancellation condition.

We conclude that this vacuum, based on an elliptically fibered Calabi-Yau
three-fold with $B=dP_{8}$, has a three family, SU(5) GUT theory supported on
the $1$-plane. This visible sector has $\alpha^{(1)}<0$ and, hence, negative
brane tension. On the other hand, the $2$-plane supports an $E_{7}$ hidden sector
which, since $\alpha^{(2)}>0$, has positive brane tension. Anomaly
cancellation requires that there be physical five-branes in the bulk, which,
since they correspond to an effective class, have positive tension.

\section*{Example 2: $F_{r}$}

In this example, we will choose the base surface of the elliptically fibered
Calabi-Yau three-fold to be
\begin{equation}
B=F_{r}.
\label{eq:84}
\end{equation}
Using (\ref{eq:39}), (\ref{eq:40}) and the fact that $c_{1}(F_{r})^{2}=8$, it
follows that, for arbitrary $r$, Eq.~(\ref{eq:56}) becomes 
\begin{equation}
c_{2}(V^{1})-\frac{1}{2}c_{2}(TX)=(\eta^{(1)}-6c_{1}(B))\sigma
-\frac{1}{2}\left(172-\frac{15}{\lambda^{(1)}}
\left(\lambda^{(1)2}-\frac{1}{4}\right)\right)F,
\label{eq:85}
\end{equation}
Eq.~(\ref{eq:62}) simplifies to
\begin{eqnarray}
\nonumber
& & c_{2}(V^{2})-\frac{1}{2}c_{2}(TX)=(\eta^{(2)}-6c_{1}(B))\sigma \\
& & \;\;\;\;\;\;\;-\frac{1}{2}\left(96-2\left(\lambda^{(2)2}
-\frac{1}{4}\right)\eta^{(2)}(\eta^{(2)}-2c_{1}(B))\right)F,
\label{eq:86}
\end{eqnarray}
and the five-brane class specified by (\ref{eq:65}) and (\ref{eq:66}) becomes
\begin{equation}
W_{B}=\sigma(12c_{1}(B)-\eta^{(1)}-\eta^{(2)})
\label{eq:87}
\end{equation}
and
\begin{equation}
a_{f}= 134
-\frac{15}{2\lambda^{(1)}}\left(\lambda^{(1)2}-\frac{1}{4}\right)
-\left(\lambda^{(2)2}-\frac{1}{4}\right)\eta^{(2)}(\eta^{(2)}-2c_{1}(B)),
\label{eq:88}
\end{equation}
where
\begin{equation}
c_{1}(F_{r})=2S+(r+2){\cal{E}}.
\label{eq:89}
\end{equation}

We now choose
\begin{equation}
\lambda^{(1)}=-\frac{3}{2}, \qquad \eta^{(1)}=2S+(r-3){\cal{E}}.
\label{eq:90}
\end{equation}
Note that this value of $\lambda^{(1)}$ satifies (\ref{eq:57}) and that
$\eta^{(1)}$ is effective for any integer $r\geq3$, which we 
henceforth assume. It is straightforward to
show that these choices satisfy the three-family condition (\ref{eq:54}) on
the visible plane. Inserting this data into (\ref{eq:85}), we conclude that
\begin{equation}
c_{2}(V^{1})-\frac{1}{2}c_{2}(TX)=-(10S+5(r+3){\cal{E}})\sigma-96F.
\label{eq:91}
\end{equation}

Furthermore, let us take
\begin{equation}
\lambda^{(2)}=1, \qquad \eta^{(2)}=6c_{1}(F_{r})
\label{eq:92}
\end{equation}
We see that this value of $\lambda^{(2)}$ satisfies (\ref{eq:63}) and that
$\eta^{(2)}=c_{1}(F_{r})+10S+5(r+2){\cal{E}}$ and, hence, satifies
(\ref{eq:64}) for even integers $r$. We hereafter assume that $r$ is even. 
Clearly $\eta^{(2)}$ is effective. Substituting this
data into (\ref{eq:86}), (\ref{eq:87}) and (\ref{eq:88}) then gives
\begin{equation}
c_{2}(V^{2})-\frac{1}{2}c_{2}(TX)=96F
\label{eq:93}
\end{equation}
and
\begin{equation}
W=(10S+5(r+3){\cal{E}})\sigma,
\label{eq:94}
\end{equation}
respectively. Note that $W$ satisfies (\ref{eq:49}) and, therefore, is an
effective class.

The tensions on the visible $1$-plane, the hidden
$2$-plane and the bulk five-brane are found by inserting these expressions in 
(\ref{eq:55}), (\ref{eq:61}) and (\ref{eq:67}) respectively. To evaluate the
signs of these tensions, one notes that
\begin{equation}
\int_{{\cal{C}}_{\omega}}{S\sigma}>0, \qquad \int_{{\cal{C}}_{\omega}}
{{\cal{E}}\sigma}>0, \qquad \int_{{\cal{C}}_{\omega}}{F}>0
\label{eq:95}
\end{equation}
since $S$, ${\cal{E}}$ and $F$ are effective classes. Inserting
Eq.~(\ref{eq:91}) into (\ref{eq:55}),  we find 
\begin{equation}
\alpha^{(1)}=-\frac{2\sqrt{2}\pi}{v^{2/3}}\left(\frac{\kappa}{4\pi}\right)^{2/3}
\int_{{\cal{C}}_{\omega}}{((10S+5(r+3){\cal{E}})\sigma+96F)}
\label{eq:96}
\end{equation}
It follows that, using (\ref{eq:95}), the tension on the visible 
$1$-plane is negative for any allowed integer $r$. Similarly, 
substituting Eq.~(\ref{eq:93}) into (\ref{eq:61}) yields
\begin{equation}
\alpha^{(2)}=\frac{2\sqrt{2}\pi}{v^{2/3}}\left(\frac{\kappa}{4\pi}\right)^{2/3}
\int_{{\cal{C}}_{\omega}}{96F}
\label{eq:97}
\end{equation}
which is positive. It follows that 
the tension on the hidden $2$-plane is positive. Finally,
putting Eq.~(\ref{eq:94}) into (\ref{eq:67}) gives
\begin{equation}
\beta=\frac{2\sqrt{2}\pi}{v^{2/3}}\left(\frac{\kappa}{4\pi}\right)^{2/3}
\int_{{\cal{C}}_{\omega}}{(10S+5(r+3){\cal{E}})\sigma}
\label{eq:98}
\end{equation}
which implies that the five-brane has positive tension. As a final check on 
this calculation, note that 
\begin{equation}
\alpha^{(1)}+\alpha^{(2)}+ \beta=0
\label{eq:99}
\end{equation}
as it must to satisfy the topological anomaly cancellation condition.

We conclude that this vacuum, based on an elliptically fibered Calabi-Yau
three-fold with $B=F_{r}$, $r\geq 4$ and even, 
has a three family, SU(5) GUT theory supported on
the $1$-plane. This visible sector has $\alpha^{(1)}<0$ and, hence, negative
brane tension. On the other hand, the $2$-plane supports an $E_{7}$ hidden sector
which, since $\alpha^{(2)}>0$, has positive brane tension. Anomaly
cancellation requires that there be physical five-branes in the bulk, which,
since they correspond to an effective class, have positive tension.

These examples establish, definitively, that $M$-theory vacua exist
in which the visible sector, hidden sector and bulk five-brane ``manifestly'' 
have negative, positive and positive tensions respectively. 
Furthermore, one can produce many examples of this type, indicating that such
vacua are by no means exceptional. Be that as it may, such 
examples are only a small subset of vacua with these properties. In this
larger class of vacua, the signs of the tensions are less manifest, and 
require further discussion. 

\section{Curve Volume and Brane Tension}

In this section, we greatly
extend the number of vacua with the property that the tensions on the visible
$1$-plane, hidden $2$-plane and bulk five-brane are negative, positive and
positive respectively. To do this, we must first
consider the relative magnitude of specific integrals over effective classes. As
discussed in \cite{RD4} and in the Appendix, the requirement that a 
holomorphic vector bundle over an elliptically fibered Calabi-Yau three-fold be
stable, restricts its K\"{a}hler class to a region near the 
boundary of the K\"{a}hler cone.
This restriction is such that the volume of any effective 
curve in the three-fold of the form $\pi^{*}\eta\cdot\sigma$, that is, 
any ``horizontal'' 
curve, is much larger than the volume of the ``vertical'' fiber
curve $F$. That is
\begin{equation}
\int_{{\cal{C}}_{\omega}}{\pi^{*}\eta\cdot\sigma} >> \int_{{\cal{C}}_{\omega}}{F}
\label{eq:100}
\end{equation}
In fact, as discussed in \cite{RD4} and in the Appendix, 
the volume of any horizontal curve can 
be made arbitrarily large relative to the volume of the fiber curve $F$. This
is accomplished by choosing the K\"{a}hler class sufficiently 
close to the K\"{a}hler 
cone boundary. This inequality will have important implications for determining
$M$-theory vacua with the required tension properties. These implications are 
most easily demonstrated with explicit examples, to which we now turn.

\section*{Example 3: $F_{r}$}

In this example, we continue to assume that equations (\ref{eq:51}) and
(\ref{eq:54}) hold, leading to a three family $SU(5)$ GUT on the visible
$1$-plane. We further maintain condition (\ref{eq:60}), which implies that the
$2$-plane supports a hidden sector with an $E_{7}$ gauge group. Choose the
base surface of the elliptically fibered three-fold to be
\begin{equation}
B=F_{r}.
\label{eq:101}
\end{equation}
The relevant equations are then (\ref{eq:85}),(\ref{eq:86}),(\ref{eq:87}) and
(\ref{eq:88}).

We now choose, as before
\begin{equation}
\lambda^{(1)}=-\frac{3}{2}, \qquad \eta^{(1)}=2S+(r-3){\cal{E}}.
\label{eq:102}
\end{equation}
This value of $\lambda^{(1)}$ satisfies (\ref{eq:57}) and $\eta^{(1)}$ is
effective for any integer $r\geq3$. These choices satisfy the three family
condition (\ref{eq:54}) on the visible plane. Inserting this data into
(\ref{eq:85}), we conclude that
\begin{equation}
c_{2}(V^{1})-\frac{1}{2}c_{2}(TX)=-(10S+5(r+3){\cal{E}})\sigma-96F.
\label{eq:103}
\end{equation}

Now, however, let us take
\begin{equation}
\lambda^{(2)}=0, \qquad \eta^{(2)}=22S+(11r+26){\cal{E}}
\label{eq:104}
\end{equation}
We see that this value of $\lambda^{(2)}$ satisfies (\ref{eq:63}) and that
$\eta^{(2)}=c_{1}(F_{r})+10c_{1}+4{\cal{E}}$ and, hence, satisfies
(\ref{eq:64}) for any integer $r$. Clearly $\eta^{(2)}$ is effective.
Substituting this data into (\ref{eq:86}), (\ref{eq:87}) and (\ref{eq:88}) 
then gives
\begin{equation}
c_{2}(V^{2})-\frac{1}{2}c_{2}(TX)=(10S+(5r+14){\cal{E}})\sigma-286F
\label{eq:105}
\end{equation}
and
\begin{equation}
W={\cal{E}}\sigma+382F,
\label{eq:106}
\end{equation}
respectively. Note that $W$ satisfies (\ref{eq:49}) and, therefore, is an
effective class.

The tensions on the visible $1$-plane, the hidden $2$-plane and the bulk
five-brane are found by inserting these expressions into (\ref{eq:55}),
(\ref{eq:61}) and (\ref{eq:67}) respectively. Inserting Eq.~(\ref{eq:103}) 
into (\ref{eq:55}), we find 
\begin{equation}
\alpha^{(1)}=-\frac{2\sqrt{2}\pi}{v^{2/3}}\left(\frac{\kappa}{4\pi}\right)^{2/3}
\int_{{\cal{C}}_{\omega}}{((10S+5(r+3){\cal{E}})\sigma+96F)}
\label{eq:107}
\end{equation}
We conclude that, using (\ref{eq:96}), the tension on the visible brane is
negative for any allowed integer $r$. Similarly, substituting Eq.~(\ref{eq:105}) into (\ref{eq:61}) yields
\begin{equation}
\alpha^{(2)}=\frac{2\sqrt{2}\pi}{v^{2/3}}\left(\frac{\kappa}{4\pi}\right)^{2/3}
\int_{{\cal{C}}_{\omega}}{((10S+(5r+14){\cal{E}})\sigma-286F)}
\label{eq:108}
\end{equation}
Here, however, we find a difference from the previous examples. This tension
is not ``manifestly'' positive because of the minus sign in the integrand.
However, it follows from (\ref{eq:100}) that
\begin{equation}
\int_{{\cal{C}}_{\omega}}{S\sigma}>>\int_{{\cal{C}}_{\omega}}{F}, \qquad
\int_{{\cal{C}}_{\omega}}{{\cal{E}}\sigma}>>\int_{{\cal{C}}_{\omega}}{F}
\label{eq:109}
\end{equation}
The K\"{a}hler class can be chosen so that, for any allowed value of $r$, the
$F$ term in (\ref{eq:108}) can be ignored relative to the $S$ and ${\cal{E}}$
terms. For the smallest value of $r$, that is, $r=3$, this can be achieved by making
the volume of the each base curve greater than eight times the volume of $F$,
which is easily done. For larger values of $r$, this constraint gets
progressively simpler to satisfy. We conclude therefore, using (\ref{eq:96}), that, 
despite the appearance of the minus sign in the integrand, the tension on the
hidden brane is positive for any allowed integer $r$. Finally, putting
Eq.~(\ref{eq:106}) into (\ref{eq:67}) gives
\begin{equation}
\beta=\frac{2\sqrt{2}\pi}{v^{2/3}}\left(\frac{\kappa}{4\pi}\right)^{2/3}\int_
{{\cal{C}}_{\omega}}{({\cal{E}}\sigma+382F)}
\label{eq:110}
\end{equation}
which implies the five-brane has positive tension. As a final check, note that 
\begin{equation}
\alpha^{(1)}+\alpha^{(2)}+\beta=0
\label{eq:111}
\end{equation}
as it must.

We conclude that this vacuum, based on an elliptically fibered Calabi-Yau
three-fold with $B=F_{r}$, $r\geq3$, has a three family, $SU(5)$ GUT supported
on the $1$-plane. This visible sector has $\alpha^{(1)}<0$ and, hence,
negative brane tension. On the other hand, the $2$-plane supports an $E_{7}$
hidden sector. The stability of the holomorphic vector bundle
$V^{2}$ is guaranteed if we choose the volume of the base curves much larger 
than the volume of the fiber curve $F$. It then follows that $\alpha^{(2)}>0$.
Hence, the hidden sector has positive brane tension. Anomaly cancellation 
requires that
there be physical five-branes in the bulk which, since they correspond to an
effective class, have positive brane tension. We remark in passing that
$\alpha^{(1)}$ is a linear function of $r$ whereas $\beta$ is independent of
$r$. This fact is physically significant, as we will discuss in the next
section.

\section*{Example 4: $F_{r}$}

In this example, we continue to assume that equations (\ref{eq:51}) and
(\ref{eq:54}) hold, leading to a three family $SU(5)$ GUT on the visible
$1$-plane. However, we now take the structure group of $V^{2}$ to be
\begin{equation}
G^{(2)}=SU(3).
\label{eq:112}
\end{equation}
That is, we choose
\begin{equation}
n^{(2)}=3.
\label{eq:113}
\end{equation}
It follows that the low energy gauge group on the $2$-plane is 
\begin{equation}
H^{(2)}=E_{6}
\label{eq:114}
\end{equation}
which is the commutant subgroup of $G^{(2)}=SU(3)$ in $E_{8}$. We need to
impose no further restrictions. Condition (\ref{eq:113}) implies that the
$2$-plane supports an $E_{6}$ supergauge theory with a certain number of
matter multiplets. That is, the $2$-plane is typical of a hidden sector. 
Note from (\ref{eq:43}) that, since we have chosen $n^{(2)}$ to be even, we
must have
\begin{equation}
\lambda^{(2)}=m^{(2)} +\frac{1}{2}
\label{eq:115}
\end{equation}
where $m^{(2)}$ is any integer. There is no further constraint on
$\eta^{(2)}$.

Since the analysis is similar to that of the previous three examples, here we
will simply state the results. We begin by choosing
\begin{equation}
\lambda^{(1)}=-\frac{3}{2}, \qquad \eta^{(1)}=2S+(r-3){\cal{E}}.
\label{eq:116}
\end{equation}
This value of $\lambda^{(1)}$ satisfies (\ref{eq:57}) and $\eta^{(1)}$ is
effective for any integer $r\geq3$. These choices satisfy the three family
condition (\ref{eq:54}) on the visible plane. With this data, we find that
\begin{equation}
c_{2}(V^{1})-\frac{1}{2}c_{2}(TX)=-(10S+5(r+3){\cal{E}})\sigma-96F.
\label{eq:117}
\end{equation}
Now take
\begin{equation}
\lambda^{(2)}=\pm\frac{1}{2}, \qquad \eta^{(2)}=22S+(11r+27){\cal{E}}.
\label{eq:118}
\end{equation}
This value of $\lambda^{(2)}$ satisfies (\ref{eq:115}) and $\eta^{(2)}$ is
effective for any allowed value of $r$. It follows from these choices that
\begin{equation}
c_{2}(V^{2})-\frac{1}{2}c_{2}(TX)=(10S+5(r+3){\cal{E}})\sigma-54F
\label{eq:119}
\end{equation}
and
\begin{equation}
W=150F.
\label{eq:120}
\end{equation}
Note that $W$ satisfies (\ref{eq:49}) and, therefore, is an effective class.

The tensions on the visible $1$-plane, the hidden $2$-plane and the bulk
five-brane are found by inserting these expressions into (\ref{eq:55}),
(\ref{eq:61}) and (\ref{eq:67}) respectively. Inserting Eq.~(\ref{eq:117})
into (\ref{eq:55}), we find 
\begin{equation}
\alpha^{(1)}=-\frac{2\sqrt{2}\pi}{v^{2/3}}\left(\frac{\kappa}{4\pi}\right)^{2/3}
\int_{{\cal{C}}_{\omega}}{((10S+5(r+3){\cal{E}})\sigma+96F)}.
\label{eq:121}
\end{equation}
It follows that, using (\ref{eq:96}), the tension on the visible brane is
negative for any allowed integer $r$. Similarly, substituting Eq.~(\ref{eq:119}) into (\ref{eq:61}) yields
\begin{equation}
\alpha^{(2)}=\frac{2\sqrt{2}\pi}{v^{2/3}}\left(\frac{\kappa}{4\pi}\right)^{2/3}
\int_{{\cal{C}}_{\omega}}{((10S+5(r+3){\cal{E}})\sigma-54F)}.
\label{eq:122}
\end{equation}
As in the previous example, this tension is not ``manifestly'' positive
because of the minus sign in the integrand. However, it follows from
(\ref{eq:109}) that, for any allowed value of $r$, the K\"{a}hler class can be chosen
so that the $F$ term in (\ref{eq:122}) can be ignored relative to the
$S$ and ${\cal{E}}$ terms. For the smallest value of $r$, that is, $r=3$, it
suffices to make the volume of each base curve greater than two times the
volume of $F$, which is easily achieved. For larger values of $r$, this
constraint gets progressively simpler to satisfy. We conclude therefore, using
(\ref{eq:96}), that, despite the appearance of the minus sign in the
integrand, the tension of the hidden brane is positive for any allowed integer
$r$. Finally, putting Eq.~(\ref{eq:120}) into (\ref{eq:67}) gives
\begin{equation}
\beta=\frac{2\sqrt{2}\pi}{v^{2/3}}\left(\frac{\kappa}{4\pi}\right)^{2/3}
\int_{{\cal{C}}_{\omega}}{150F}
\label{eq:123}
\end{equation}
which implies that the five-brane has positive tension. As a final check, note
that
\begin{equation}
\alpha^{(1)}+\alpha^{(2)}+\beta=0
\label{eq:124}
\end{equation}
as it must.

We conclude that this vacuum, based on an elliptically fibered Calabi-Yau
three-fold with $B=F_{r}$, $r\geq3$, has a three family, $SU(5)$ GUT supported
on the $1$-plane. This visible sector has $\alpha^{(1)}<0$ and, hence,
negative brane tension. On the other hand, the $2$-plane supports an $E_{6}$
hidden sector. Since the stability of the holomorphic vector bundle
$V^{2}$ requires that we choose the volume of the base curves much larger than
the volume of the fiber curve $F$, it follows that $\alpha^{(2)}>0$. Hence,
the hidden sector has positive brane tension. Anomaly cancellation requires
that there be physical five-branes in the bulk which, since they correspond to
an effective class, have positive brane tension. We remark in passing that,
unlike $\alpha^{(1)}$, $\beta$ depends only on the small volume fiber curve
$F$. This fact is physically relevant, as we will discuss in the next section.

These examples establish, definitively, that there is an enormous class of
$M$-theory vacua which have the property that the tensions on the visible
sector, hidden sector and bulk five-brane are negative, positive and positive
respectively.

\section{The Ratio $\beta/|\alpha^{(1)}|$}

In the recently proposed Ekpyrotic theory of cosmology 
\cite{EU}, a fundamental input
parameter is the ratio of the five-brane tension $\beta$ to the magnitude of
the visible brane tension $|\alpha^{(1)}|$. In this section, we examine the
values of that ratio in each of the representative examples presented above.

\subsection*{Example 1:}

Choosing the K\"{a}hler class so that 
\begin{equation}
\int_{{\cal{C}}_{\omega}}{c_{1}(dP_{8})\sigma} >>\int_{{\cal{C}}_{\omega}}{F}
\label{eq:125}
\end{equation}
allows one to safely ignore the contribution of the fiber curve to the
tensions of both the bulk five-brane and the visible $1$-plane. It then
follows from Eqs.~(\ref{eq:80}) and (\ref{eq:82}) that
\begin{equation}
\beta/|\alpha^{(1)}|\sim \frac{3}{4}.
\label{eq:126}
\end{equation}

\subsection*{Example 2:}

Choosing the K\"{a}hler class so that inequalities (\ref{eq:109}) are satisfied,
allows one to disregard the contribution of the fiber curve to the
tensions. Then, we find from equations (\ref{eq:96}) and (\ref{eq:98}) that
\begin{equation}
\beta/|\alpha^{(1)}|\sim 1.
\label{eq:128}
\end{equation}

Note that the $\beta/|\alpha^{(1)}|$ ratio in both of these examples is near
unity. It follows that neither of these vacua would be suitable candidates for
the Ekpyrotic cosmological scenario, at least not as presently formulated
in \cite{EU}. However, as we now show, this is only true of a restricted 
set of $M$-theory vacua.

\subsection*{Example 3:}

Choosing the K\"{a}hler class so that inequalities (\ref{eq:109}) are satisfied,
again allows one to safely ignore the contribution of the fiber curve to the
tensions. Then, from equations (\ref{eq:107}) and (\ref{eq:110}) we find that
\begin{equation}
\beta/|\alpha^{(1)}| < \frac{1}{5(r+3)}
\label{eq:129}
\end{equation}
It is important to note that the value of integer $r$ for Hirzebruch surfaces
$F_{r}$ is unrestricted, and can be made as large as desired. It follows that,
for large integer $r$, the $\beta/|\alpha^{(1)}|$ ratio can be made arbitrarily 
small.  Such vacua are perfectly suited to support the Ekpyrotic cosmology
presented in \cite{EU}.

\subsection*{Example 4:}

This example differs somewhat from the previous three in that $\beta$ depends
only on the volume of the fiber curve. It follows from equations (\ref{eq:109}),
(\ref{eq:121}) and (\ref{eq:123}) that
\begin{equation}
\beta/|\alpha^{(1)}| < \left(\frac{30}{r+3}\right) \frac{\int_{{\cal{C}}_{\omega}}{F}}
{\int_{{\cal{C}}_{\omega}}{{\cal{E}}\sigma}}
\label{eq:130}
\end{equation}
For any fixed value of integer $r$, the $\beta/|\alpha^{(1)}|$ 
ratio can be made as small as desired by appropriate choice of the K\"{a}hler
structure. Clearly, in this type of vacua, the $\beta/|\alpha^{(1)}|$ ratio 
is naturally very small. Such vacua have all the properties required by 
the Ekpyrotic cosmological scenario.

\section{Conclusion}

We have shown that there are classes of BPS vacua in
heterotic $M$-theory that exhibit negative tension on the visible orbifold
plane, positive tension on the hidden plane and positive tension on the
five-branes in the bulk. We chose as background geometries 
elliptically fibered Calabi-Yau three-folds with certain del Pezzo and 
Hirzebruch bases. However, it is straightforward to extend our results to any 
elliptically fibered Calabi-Yau three-fold. Furthermore, in our examples,
we chose the low energy gauge group on the visible brane to be
$G^{(1)}=SU(5)$. Similar conclusions will be reached, however, if one takes
other physically relevant gauge groups, such as $G^{(1)}=E_{6}, SO(10)$ or the
standard model gauge group $SU(3)_{C} \times SU(2)_{L} \times U(1)_{Y}$.
In the same way, although we chose the gauge group on the hidden
brane to be $G^{(2)}=E_{7}$ and $E_{6}$, many other choices, such as of
$G^{(2)}=SO(10)$, will lead to similar results. That is, the class of 
BPS solutions in $M$-theory with
negative tension on the visible brane and positive tensions on the hidden
brane and bulk five-branes is large and robust. Although it is hard to
quantify, in practice it is as easy to find $M$-theory vacua with this property as
to find vacua with the reverse property, that is,  positive tension on the visible
brane.

In addition, we examined the ratio, $\beta/|\alpha^{(1)}|$, of the tension of
the bulk brane to the visible brane tension in vacua of this type. 
We found that, for many vacua, this
ratio is of order unity. However, it was easy to construct examples in
which this ratio is small. In fact, there are several different classes of
physically sensible vacua for which this ratio is arbitrarily small.

We conclude that the properties of vacua required in the examples
of Ekpyrotic cosmology presented in \cite{EU}, that is, a negative tension 
visible brane, a positive tension hidden brane, positive tension bulk five-branes
and a small $\beta/|\alpha^{(1)}|$ ratio, are found among large classes of physically
realistic vacua of heterotic $M$-theory.

\bigskip
\bigskip
\section*{Acknowledgments}
We would like to thank Andre Lukas and Dan Waldram for helpful conversations.
This work was supported in part by the National Science Foundation grant
DMS-9802456 (RYD), the Natural Sciences and 
Engineering Research Council of Canada (JK),
the US Department of Energy grants DE-FG02-91ER40671 (JK and PJS), 
DE-FG02-96ER40959 and DE-AC02-76-03071 (BAO), and by PPARC-UK (NT). 
\bigskip

\section*{Appendix}

In order to preserve $N=1$ supersymmetry, the vector bundles
$V^{n}$ must be stable. (Actually, it suffices for them to be
poly-stable. Poly-stability is a property intermediate between stability
and semi-stability. All the bundles we construct in this paper
are actually stable, so we do not need to worry about poly-stability here.)

It is important to note that the notion of stability of a vector bundle
$V$ depends on the choice of the K\"{a}hler class $\omega$ on the elliptically
fibered Calabi-Yau
three-fold $X$. The set of all possible K\"{a}hler classes forms the K\"{a}hler cone. 
This is an open, convex cone in the vector space $H^2(X,{\bf R})$, 
given by the condition that the integral of ${\omega}^i$ on each effective 
$i$-dimensional complex subspace of $X$ is positive. In particular, 
the integral of $\omega$ itself on each effective curve must be positive. This
fact was used in equations (\ref{eq:79}) and (\ref{eq:95}) in the text.

It is known that as $\omega$ varies in the K\"{a}hler cone, the moduli space of stable
bundles can jump around as a result of bundles which are stable with
respect to one K\"{a}hler class becoming unstable with respect to another.
Typically, the K\"{a}hler cone is divided by walls into open chambers; the
moduli space is independent of $\omega$ in the interior of each chamber, but
undergoes a flop-like transition when a wall is crossed.

Now the spectral construction \cite{FMW1,AJ} produces bundles $V$ on $X$ with the
property that the restriction $V {\vert}_F$ to a general elliptic fiber
$F$ is semi-stable. (But not stable: typically $V {\vert}_F$ is 
either a sum of several line bundles, all of the same slope, or an
extension involving such line bundles.) Let $w$ be a K\"{a}hler class on the
base $B$. Then $w^2$ is some positive multiple of the class of a point
in $B$. Therefore, for the pullback ${\pi}^*w$ on $X$, we have that
$({\pi}^*w)^2$ is some positive multiple of the class of the fiber $F$. The
semi-stability of $V {\vert}_F$, therefore, amounts to saying that $V$ is
``semi-stable with respect to ${\pi}^*w$''.

This result is incomplete on two counts. First, we need to strengthen
semi-stability to stability. But more importantly, ${\pi}^*w$ is not
a K\"{a}hler class on $X$. Its integral on each effective curve $C$ in $X$
is indeed non-negative: it is the same as the integral of the original
K\"{a}hler class $w$ on the image $\pi(C)$ in $B$. However, there is one
curve, namely the fiber $F$, for which the image $\pi(C)$ and, hence, also 
the integral, vanish. This means that ${\pi}^*w$ is on the boundary of
the K\"{a}hler cone. We must move to the interior in order for stability of
$V$ to make sense.

It turns out that we can indeed cure both problems simultaneously by 
moving ever so slightly into the interior of the K\"{a}hler cone, specifically 
into the chamber which is closest to the 
given boundary point ${\pi}^*w$. To do this,
choose any K\"{a}hler class $\omega_0$ on $X$, and consider the combination 
$$ \omega = {\pi}^*w + {\epsilon} \omega_0, \qquad {\epsilon} > 0. $$
$\omega$, unlike ${\pi}^*w$, is a K\"{a}hler class on $X$. 
It has the property that for any
effective class $c$ in $H^2(B,{\bf Z})$, the volume of 
${\sigma} \cdot {\pi}^*c$ with respect to $\omega$ on $X$ 
is at least the volume of $c$ with respect to the
K\"{a}hler class $w$ on the base $B$. On the other hand, the volume of $F$ 
with respect to $\omega$ is $\epsilon \int_{{\cal{C}}_{\omega_0}}{F}$. 
By taking $\epsilon$
sufficiently small, we can thus make the volume of $F$ arbitrarily small
compared to the volume of effective classes associated with the base. This result
was used in equations (\ref{eq:100}), (\ref{eq:109}) and (\ref{eq:125}) in the
text.

The crucial property of this K\"{a}hler class $\omega$ is that, for
$\epsilon$ sufficiently small, a vector bundle $V$ which arises by the spectral
construction from an irreducible spectral cover is actually stable with
respect to this choice of $\omega$. This was proven in Theorem 7.1 of
\cite{FMW2}. It has been used in many of the heterotic $M$-theory constructions 
including \cite{RD3}, where it was employed to reduce the GUT gauge 
group $SU(5)$ to the standard model gauge group $SU(3)_{C} \times SU(2)_{L} 
\times U(1)_{Y}$. (The stability 
of $V$ is discussed in section (5.2) of that reference, and the specific 
K\"{a}hler class $\omega$ used there is given in (6.1.3).) 
In the present situation, this implies that the bundle $V$ can be taken 
to be stable with respect to a K\"{a}hler class $\omega$ while keeping the volume
of the fiber $F$ arbitrarily small compared to the volumes of effective curves
associated with the base.

Let us explain why $V$ is stable with respect to $\omega$. Stability
means that the slope $\mu_\omega(W)$ of any proper sub-bundle (or sub-sheaf) 
$W \subset V$ is strictly less than $\mu_\omega(V)$. Here the slope (with
respect to $\omega$) is defined as 
$\mu_\omega(W) = \frac{c_1(W) \cdot \omega^2}{rank(W)}$.
Even though ${\pi}^*w$ is in the boundary of the K\"{a}hler cone, and so
is not a K\"{a}hler class on $X$, we can still define the slope 
$\mu_{{\pi}^*w}(W)= \frac{c_1(W) \cdot (\pi^*w)^2}{rank(W)}$ 
with respect to it. Since $({\pi}^*w)^2$
is some positive multiple of the class of the fiber $F$, the resulting notion
of ``semi-stability with respect to ${\pi}^*w$'' is implied by
semi-stability of the restrictions $V {\vert}_F$ to the fibers.
Assume that $V$ is not stable with respect to $\omega$, then there is a
``destabilizing'' sub-bundle $W \subset V$ with $\mu_\omega(W) \ge \mu_\omega(V)$.
But semi-stability along the fibers says that $\mu_{{\pi}^*w}(W) \le
\mu_{{\pi}^*w}(V)$. If we had equality, it would follow that $W$
arises by the spectral construction from a proper sub-variety of the
spectral cover of $V$, contradicting the assumption that this cover is
irreducible. So we must have a strict inequality $\mu_{{\pi}^*w}(W)
<\mu_{{\pi}^*w}(V)$.
But then it is easy to see that, by taking $\epsilon$ small enough, we can
also ensure that $\mu_{\omega}(W) < \mu_{\omega}(V)$, so $W$ cannot destabilize $V$
after all. (For more details, see \cite{FMW2}.)

\end{document}